\documentstyle[twoside]{article}

\input{psfig} 

\def\kmu{$K^+\to \mu^+ \pi^0 \nu$}

\def\rkm{$K^+\to \mu^+ \nu \gamma$}

\def\kpp{$K^+\to \pi^+ \pi^+ \pi^-$}

\catcode`\@=11
\long\def\@makefntext#1{
\protect\noindent \hbox to 3.2pt {\hskip-.9pt  
$^{{\eightrm\@thefnmark}}$\hfil}#1\hfill}		%CAN BE USED 

\def\thefootnote{\fnsymbol{footnote}}
\def\@makefnmark{\hbox to 0pt{$^{\@thefnmark}$\hss}}	%ORIGINAL 
	
\def\ps@myheadings{\let\@mkboth\@gobbletwo
\def\@oddhead{\hbox{}
\rightmark\hfil\eightrm\thepage}   
\def\@oddfoot{}\def\@evenhead{\eightrm\thepage\hfil
\leftmark\hbox{}}\def\@evenfoot{}
\def\sectionmark##1{}\def\subsectionmark##1{}}

\oddsidemargin=\evensidemargin
\renewcommand{\thefootnote}{\fnsymbol{footnote}}

\newcounter{sectionc}\newcounter{subsectionc}\newcounter{subsubsectionc}
\renewcommand{\section}[1] {\vspace{12pt}\addtocounter{sectionc}{1} 
\setcounter{subsectionc}{0}\setcounter{subsubsectionc}{0}\noindent 
	{\tenbf\thesectionc. #1}\par\vspace{5pt}}
\renewcommand{\subsection}[1] {\vspace{12pt}\addtocounter{subsectionc}{1} 
	\setcounter{subsubsectionc}{0}\noindent 
	{\bf\thesectionc.\thesubsectionc. {\kern1pt \bfit #1}}\par\vspace{5pt}}
\renewcommand{\subsubsection}[1] {\vspace{12pt}\addtocounter{subsubsectionc}{1}
	\noindent{\tenrm\thesectionc.\thesubsectionc.\thesubsubsectionc.
	{\kern1pt \tenit #1}}\par\vspace{5pt}}

\newcounter{appendixc}
\newcounter{subappendixc}[appendixc]
\newcounter{subsubappendixc}[subappendixc]
\renewcommand{\thesubappendixc}{\Alph{appendixc}.\arabic{subappendixc}}
\renewcommand{\thesubsubappendixc}
	{\Alph{appendixc}.\arabic{subappendixc}.\arabic{subsubappendixc}}

\renewcommand{\appendix}[1] {\vspace{12pt}
        \refstepcounter{appendixc}
        \setcounter{figure}{0}
        \setcounter{table}{0}
        \setcounter{lemma}{0}
        \setcounter{theorem}{0}
        \setcounter{corollary}{0}
        \setcounter{definition}{0}
        \setcounter{equation}{0}
        \renewcommand{\thefigure}{\Alph{appendixc}.\arabic{figure}}
        \renewcommand{\thetable}{\Alph{appendixc}.\arabic{table}}
        \renewcommand{\theappendixc}{\Alph{appendixc}}
        \renewcommand{\thelemma}{\Alph{appendixc}.\arabic{lemma}}
        \renewcommand{\thetheorem}{\Alph{appendixc}.\arabic{theorem}}
        \renewcommand{\thedefinition}{\Alph{appendixc}.\arabic{definition}}
        \renewcommand{\thecorollary}{\Alph{appendixc}.\arabic{corollary}}
        \renewcommand{\theequation}{\Alph{appendixc}.\arabic{equation}}
        \noindent{\tenbf Appendix \theappendixc #1}\par\vspace{5pt}}
\newcommand{\subappendix}[1] {\vspace{12pt}
        \refstepcounter{subappendixc}
        \noindent{\bf Appendix \thesubappendixc. {\kern1pt \bfit #1}}
	\par\vspace{5pt}}
\newcommand{\subsubappendix}[1] {\vspace{12pt}
        \refstepcounter{subsubappendixc}
        \noindent{\rm Appendix \thesubsubappendixc. {\kern1pt \tenit #1}}
	\par\vspace{5pt}}

\newcommand{\textlineskip}{\baselineskip=13pt}
\newcommand{\smalllineskip}{\baselineskip=10pt}

\def\eightcirc{
\begin{picture}(0,0)
\put(4.4,1.8){\circle{6.5}}
\end{picture}}
\def\eightcopyright{\eightcirc\kern2.7pt\hbox{\eightrm c}} 

\newcommand{\copyrightheading}[1]
	{\vspace*{-2.5cm}\smalllineskip{\flushleft
	{\footnotesize International Journal of Modern Physics A, #1}\\
	{\footnotesize $\eightcopyright$\, World Scientific Publishing
	 Company}\\
	 }}

\newcommand{\publisher}[2]{{\begin{center}\footnotesize\smalllineskip 
	Received #1\\
	Revised #2
	\end{center}
	}}

\def\abstracts#1#2#3{{
	\centering{\begin{minipage}{4.5in}\baselineskip=10pt\footnotesize
	\parindent=0pt #1\par 
	\parindent=15pt #2\par
	\parindent=15pt #3
	\end{minipage}}\par}} 

\renewenvironment{thebibliography}[1]
	{\frenchspacing
	 \ninerm\baselineskip=11pt
	 \begin{list}{\arabic{enumi}.}
	{\usecounter{enumi}\setlength{\parsep}{0pt}
	 \setlength{\leftmargin 12.7pt}{\rightmargin 0pt} %FOR 1--9 ITEMS
	 \setlength{\itemsep}{0pt} \settowidth
	{\labelwidth}{#1.}\sloppy}}{\end{list}}

\newcounter{itemlistc}
\newcounter{romanlistc}
\newcounter{alphlistc}
\newcounter{arabiclistc}

\newcommand{\fcaption}[1]{
        \refstepcounter{figure}
        \setbox\@tempboxa = \hbox{\footnotesize Fig.~\thefigure. #1}
        \ifdim \wd\@tempboxa > 5in
           {\begin{center}
        \parbox{5in}{\footnotesize\smalllineskip Fig.~\thefigure. #1}
            \end{center}}
        \else
             {\begin{center}
             {\footnotesize Fig.~\thefigure. #1}
              \end{center}}
        \fi}

\newcommand{\tcaption}[1]{
        \refstepcounter{table}
        \setbox\@tempboxa = \hbox{\footnotesize Table~\thetable. #1}
        \ifdim \wd\@tempboxa > 5in
           {\begin{center}
        \parbox{5in}{\footnotesize\smalllineskip Table~\thetable. #1}
            \end{center}}
        \else
             {\begin{center}
             {\footnotesize Table~\thetable. #1}
              \end{center}}
        \fi}

\def\@citex[#1]#2{\if@filesw\immediate\write\@auxout
	{\string\citation{#2}}\fi
\def\@citea{}\@cite{\@for\@citeb:=#2\do
	{\@citea\def\@citea{,}\@ifundefined
	{b@\@citeb}{{\bf ?}\@warning
	{Citation `\@citeb' on page \thepage \space undefined}}
	{\csname b@\@citeb\endcsname}}}{#1}}

\newif\if@cghi
\def\cite{\@cghitrue\@ifnextchar [{\@tempswatrue
	\@citex}{\@tempswafalse\@citex[]}}
\def\citelow{\@cghifalse\@ifnextchar [{\@tempswatrue
	\@citex}{\@tempswafalse\@citex[]}}
\def\@cite#1#2{{$\null^{#1}$\if@tempswa\typeout
	{IJCGA warning: optional citation argument 
	ignored: `#2'} \fi}}

\def\pmb#1{\setbox0=\hbox{#1}
	\kern-.025em\copy0\kern-\wd0
	\kern.05em\copy0\kern-\wd0
	\kern-.025em\raise.0433em\box0}

\def\fnt#1#2{\footnotetext{\kern-.3em
	{$^{\mbox{\scriptsize #1}}$}{#2}}}

\def\fpage#1{\begingroup
\voffset=.3in
\thispagestyle{empty}\begin{table}[b]\centerline{\footnotesize #1}
	\end{table}\endgroup}

\def\runninghead#1#2{\pagestyle{myheadings}
\markboth{{\protect\footnotesize\it{\quad #1}}\hfill}
{\hfill{\protect\footnotesize\it{#2\quad}}}}
\font\tenrm=cmr10
\font\tenit=cmti10 
\font\tenbf=cmbx10
\font\bfit=cmbxti10 at 10pt
\font\ninerm=cmr9

\font\eightrm=cmr8

\def\qed{\hbox{${\vcenter{\vbox{			%HOLLOW SQUARE
   \hrule height 0.4pt\hbox{\vrule width 0.4pt height 6pt
   \kern5pt\vrule width 0.4pt}\hrule height 0.4pt}}}$}}

\renewcommand{\thefootnote}{\fnsymbol{footnote}}	%USE SYMBOLIC FOOTNOTE

\begin{document}

\runninghead{Tests of the Standard Model using Muon 
Polarization Asymmetries in Kaon Decay
 $\ldots$} {Tests of the Standard Model using Muon 
Polarization Asymmetries in Kaon Decay
 $\ldots$}

\normalsize\textlineskip
\thispagestyle{empty}
\setcounter{page}{1}

\copyrightheading{}			%{Vol. 0, No. 0 (1993) 000--000}

\vspace*{0.88truein}

\fpage{1}

\centerline{\bf  TESTS OF THE STANDARD MODEL USING MUON }
\vspace*{0.035truein}
\centerline{\bf POLARIZATION ASYMMETRIES IN KAON DECAYS}
\vspace*{0.37truein}
\centerline{\footnotesize MILIND V. DIWAN}
\vspace*{0.015truein}
\centerline{\footnotesize\it Physics Department, 
Brookhaven National Laboratory, Upton, NY 11973}
\baselineskip=10pt
\vspace*{10pt}
\centerline{\footnotesize HONG MA}
\vspace*{0.015truein}
\centerline{\footnotesize\it Physics Department,
Brookhaven National Laboratory, Upton, NY, 11973}
\baselineskip=10pt
\vspace*{0.225truein}
%\centerline{Submitted to the International Journal of Modern Physics A}
%\centerline{22 November 2000}
\publisher{(22 November 2000)}{(4 January 2001)}

\textlineskip

\vspace*{0.21truein}
\abstracts{
We have examined the physics and the experimental feasibility 
of studying various kaon decay processes in which the polarization of a
muon in the final state is measured.  Valuable information on CP
violation, the quark mixing (CKM) matrix, 
and new physics can be obtained from such
measurements.  
We have considered muon polarization  in $K_L \to \mu^+ \mu^-$ and
$K  \to \pi  \mu^+ \mu^-$ decays. Although the effects are small, or difficult to 
measure because of the small branching ratios involved, these 
studies could provide clean measurements of 
the CKM parameters. 
The experimental difficulty appears comparable to 
the observation of $K \to \pi \nu \bar \nu$. 
New sources of physics, involving non-standard CP violation,  could 
produce effects observable in these
measurements. 
%In particular,
%models of non-standard CP violation that produce the baryon asymmetry
%of the universe could also produce effects observable in these
%measurements. 
Limits from new results on  the neutron and
electron electric dipole moment, and $\epsilon^\prime \over \epsilon$
in neutral kaon decays, do not eliminate certain  models that could 
contribute to the signal.   A detailed examination of 
 muon polarization out of the decay plane 
in \kmu ~and \rkm ~decays also appears to be of interest.
  With current 
  kaon beams and detector techniques, it is possible to
measure the T-violating
 polarization for \kmu ~with uncertainties  approaching $\sim 10^{-4}$. 
This level of  sensitivity would provide an interesting
 probe of new physics.}{}{}

%\textlineskip			%) USE THIS MEASUREMENT WHEN THERE IS
%\vspace*{12pt}			%) NO SECTION HEADING

%\vspace*{1pt}\textlineskip	%) USE THIS MEASUREMENT WHEN THERE IS
%\section{General Appearance}	%) A SECTION HEADING
%\vspace*{-0.5pt}
%\noindent

\textheight=7.8truein
\setcounter{footnote}{0}
\renewcommand{\thefootnote}{\alph{footnote}}

\section{Introduction}

\noindent 
We have examined the possibility of measuring muon polarization 
 asymmetries that are sensitive to 
P, T or CP symmetries; these are 
tabulated in Table \ref{list1}. 
Observation of the possible effects requires 
high fluxes ($> 10^{12}$ K decays per year)
 of kaons,  
 now available at several accelerator facilities,
notably at  Brookhaven National Laboratory AGS and at the KEK-PS.
In the near future,  the Fermilab main injector, as well 
as the Japanese Hadron Factory, could deliver much higher intensities of 
separated as well as unseparated kaon beams.   The $\phi$ meson factory, 
DAPHNE, % KLOE, 
is also a new facility with an intense pure-kaon flux.
Thus far,  
CP violation has been  observed conclusively only
 in  the neutral kaon system.
Although a  theoretical description of  
CP-violation in the neutral kaon system is available in the 
single complex phase 
of the standard-model Cabibbo-Kobayashi-Maskawa  (CKM)
 matrix,  part of, or the entire  
phase,  can have origin in 
deeper causes that have so far eluded experimental scrutiny.
During this    past decade,  experiments 
at FNAL and at CERN,   focusing on the measurement of the direct 
$K^0_L \to \pi \pi$ transition,  or $\epsilon^\prime \over \epsilon$, have 
reported ever-improved 
 results, the latest being
 $\mbox{Re}(\epsilon'/\epsilon) = (28.0\pm 3.0\pm 2.8) \times 10^{-4}$
(FNAL$^{\ref{ktev}}$) and  
$\mbox{Re}(\epsilon'/\epsilon) =  (14.0 \pm 4.3 ) \times 10^{-4}$ (CERN$^{\ref{na48}}$).
These provide  conclusive evidence for the presence of  direct CP violation in 
$K^0$ decays, and  
there is great 
theoretical effort in progress  to interpret
 these numbers.$^{\ref{rosner}-\ref{winwolf}}$
We will not attempt here to review $\epsilon'/\epsilon$; but it is clear that
it is not yet certain if the standard model description of CP violation 
can accommodate  these results. In addition to 
$\epsilon'/\epsilon$, rare kaon decays 
are also of interest for understanding the CKM matrix. For a recent 
 review of rare kaon 
decay processes see Ref. \ref{skrev}.

%been inconclusive in  revealing the true nature of CP-violation. 
%
% Need update   e' measurement here with new numbers and 
% implication 
%
 Over the next decade, ambitious efforts towards gaining a better 
 understanding CP-violation 
and the CKM matrix elements will be pursued at  B-factories.
 The importance of these efforts is 
undeniable, but it is also worthwhile 
 to investigate the possibility that some or 
all of the 
CP-violation arises from effects outside of the minimal standard model,  
particularly outside of the current CKM matrix.  

It should be recalled that  CP-violation is required to generate  
the observed baryon asymmetry in the universe, and it 
is now accepted that  the CP-violation embodied in the CKM 
matrix does not have sufficient strength to serve this 
purpose.$^{\ref{mclerran}}$ 
It is important to 
examine if sources of physics beyond the standard model 
that could generate the baryon 
asymmetry   can also generate 
CP or T violating muon polarizations 
in the kaon decay modes given in Table \ref{list1}. 

\begin{table}
\begin{center}
\begin{tabular}{clccc}\hline
\hline
 & Decay                &  Correlations & Symmetries \\
  &                     &              &  tested    \\
\hline
(1) & $K^+\to \pi^0 \mu^+ \nu$   &  $\vec s_\mu\cdot (\vec p_\mu\times \vec p_\pi)$ & T \\
\hline
(2) & $K^+\to  \mu^+ \nu \gamma$  &  $\vec s_\mu\cdot (\vec p_\mu\times \vec p_\gamma)$ & T \\
\hline
(3) &  $K_L\to  \mu^+ \mu^-$ &   $\vec s_\mu\cdot \vec p_\mu$ & P, CP \\
\hline
(4) &  $K^+ \to \pi^+  \mu^+ \mu^-$ &  $\vec s_\mu\cdot \vec p_\mu$ & P \\
(5) &                             & $\vec s_\mu\cdot (\vec p_{\mu^+}\times \vec p_{\mu^-})$ & T \\
(6) &                             & $(\vec s_{\mu^\pm} \cdot \vec p_{\mu^\pm}) \vec s_{\mu^\mp} \cdot (\vec p_{\mu^+}\times \vec p_{\mu^-})$ & P, T \\
\hline 
(7) &  $K^0_L \to \pi^0  \mu^+ \mu^-$ &  $\vec s_\mu\cdot \vec p_\mu$ & P \\
(8) &                             & $\vec s_\mu\cdot (\vec p_{\mu^+}\times \vec p_{\mu^-})$ & T \\
\hline
\hline 
\end{tabular}
\end{center}
\caption{\sl Decay modes and  polarization asymmetries 
(or correlations) of interest in K decays. 
The symbols $\vec s$ and $\vec p$ refer to the spin and momentum 
vectors in the decays.
 }
\label{list1}
\end{table}

\begin{table}
\begin{center}
\begin{tabular}{llccccc}
\hline
\hline
 & Mode & Branch. &  Standard   & Final      &  Non-SM    & Ref. \\
      &      & Fraction  &  Model    & State Int. &  value      &      \\
\hline 
(1) & $K^+\to \pi^0 \mu^+ \nu$ & 0.032 & 0.0 & $\sim 10^{-6}$ & $\le 10^{-3}$ & [\ref{garisto},\ref{belanger}] \\
\hline
(2) & $K^+\to \mu^+ \nu \gamma$ & $5.5\times 10^{-3}$ & 0.0 & $\sim 10^{-3}$ & $\le 10^{-3}$ & [\ref{kobayashi}] \\
\hline
(3) & $K_L \to \mu^+ \mu^-$ & $7.2\times 10^{-9}$ & $\sim 0.002$ & 0.0 & $\le 10^{-2}$ & [\ref{botella}, \ref{wolf1}] \\
\hline
(4) &   $K^+\to \pi^+ \mu^+ \mu^-$ & $7.6\times 10^{-8}$  &  $\sim 10^{-2}$ & -- & -- & [\ref{wise1} --\ref{bgt}] \\
(5) &                &  & 0.0 & $\sim 10^{-3}$  & $\sim 10^{-3}$ & [\ref{anbg}] \\
(6) &               &   & $\sim 0.06$ & $\sim 0.0$ & $\sim 0.1$ & [\ref{anbg}] \\
\hline 
(7) &   $K^0_L\to \pi^0 \mu^+ \mu^-$ &  $\sim 5 \times 10^{-12}$  &  -- & -- & -- & [\ref{pi0mumu},\ref{ktevpmm}] \\
(8) &                &  & $\sim 0.5$ &  --   &  --  & [\ref{pi0mumu}] \\
\hline
\end{tabular}
\end{center}
\caption{\sl The decay modes and asymmetries of interest; 
the row numbers correspond to those in Table \ref{list1}.
The other columns are:
the known standard-model branching ratio (note that the 
 $K_L \to \pi^0 \mu^+ \mu^-$ branching ratio is not yet measured, the 
present 90\% C.L. limit is $< 3.8\times 10^{-10}$),
 the estimated standard-model value of the asymmetry,
the value due to final-state interactions,
 the maximum possible
value allowed by non-standard  
physics, and the theoretical reference. 
Some of the results 
have been adjusted to account for more recent values  of the top
quark mass (174 $GeV/c^2$). In the case of $K_L\to \mu^+ \mu^-$ 
and $K^+ \to \pi^+ \mu^+ \mu^-$,
the theoretical estimates for  maximum possible 
non-standard 
contributions to T violation do not agree; we chose the mean value of
the different estimates. The ``--'' means that there is no reliable prediction. 
}
\label{list2}
\end{table}

We will now briefly consider the measurement of muon polarization. 
This has been discussed before in the literature.$^{\ref{commins},\ref{vernon}}$
Muon polarization in the experiments under consideration
would be observed  by stopping the muons in some appropriate 
material,
 and measuring the direction of the decay electron (positron in the case of $\mu^+$).
The muon decay spectrum is given by: 
\begin{eqnarray} 
{dN\over dz d\Omega} = {z^2\over 2 \pi}\left( (3-2z) \pm |\vec P| \cos\theta (2 z -1 )\right)  
\label{mudecay}
\end{eqnarray} 
where the positive sign in the brackets is for $\mu^+$ decay and the negative sign is for $\mu^-$ decay, 
$z = 2 E_e/m_\mu$, and $\theta$ is the angle between the polarization $\vec{P}$ and the direction of 
the  positron or electron.  
We will now restrict ourselves to only $\mu^+$ decays. 
The magnitude of the polarization $\vec P$,  multiplied by  a dilution factor $D(z_0)$, 
  is  the 
asymmetry $A$ in the number of decays  that produce  positrons 
forwards  versus backwards with respect to 
a plane normal to the polarization vector.
\begin{eqnarray}
A = {N_1 - N_2 \over N_1 + N_2}  = D(z_0) |\vec P| 
\label{ass1f}
\end{eqnarray}
where $N_1$ and $N_2$ are the number of forward and backward decays, respectively.
$D(z_0)$ is the asymmetry dilution factor that depends on 
 the lower-energy cutoff 
$z_0 = 2 E_0/m_\mu$, where $E_0$ is the minimum observable energy of the electron
from $\mu$ decays.
 The uncertainty on the asymmetry is given by $\delta A = \sqrt{1-A^2\over N}$,
where $N=N_1 + N_2$.  When the asymmetry is small, the error on 
 the polarization is given by $\delta P = {1\over D(z_0) \sqrt{N}}$. 
In practical devices, there is usually a lower 
cutoff on the positron energy. 
Integrated 
over the entire spectrum,  the asymmetry is $|\vec P|/3$.
Clearly, both the total number of detected decays $N$, and the dilution factor $D$, depend on 
the low-energy cutoff. Best performance is reached when the lower cutoff is 
at about $z_0\sim 0.75$, which keeps approximately 1/2 of the spectrum and 
corresponds to a 
%has an asymmetry 
dilution factor of  $\sim 0.5$.  Often, the muons are precessed by a small magnetic field 
(at a rate of 42.5 kHz per Gauss)
perpendicular to the direction of the spin. This makes it possible to measure  
two components of the polarization, as well as eliminate  systematic differences
in detector efficiencies, at only a cost of $\sqrt{2}$ loss in the dilution 
factor.

Other considerations that determine the asymmetry are 
depolarization of the stopped $\mu^+$, and confusion from the 
multiple scattering of the decay positron.   It is well known that multiple 
Coulomb scattering of the muon, as it comes to rest, does not cause depolarization;
however, after it slows down to atomic velocities, there are many
processes that can lead to 
depolarization. The amount of
depolarization is strongly material dependent.$^{\ref{vernon}}$ 
Carbon (in graphite form) and aluminum are considered good materials for 
muon polarimeters because they preserve the polarization. Detectors such as wire chambers 
or scintillators must be placed next to blocks of graphite or aluminum to detect the 
decay positrons. No practical active materials such as plastic 
scintillator or scintillating 
crystals have been found  that would preserve polarization. Dilution of the 
asymmetry  from multiple scattering of the positron depends on the geometry 
and the energy cutoff.  Ultimately,  the  dilution factor  must be measured 
in each experiment to determine the sensitivity of the result.

\section{$K^+\to \pi^0 \mu^+ \nu$} 

\noindent 
The transverse, or  out-of-plane, muon polarization in this 
 decay has  been  analyzed previously.$^{\ref{garisto},
\ref{belanger}}$ The decay amplitude, $\cal M$, can be written  
as follows:
\begin{eqnarray}
{\cal M}  = {G_F\over 2} \sin\theta_c f_+(q^2) \left((p_K+p_\pi)^\lambda+\xi(q^2)
(p_K-p_\pi)^\lambda)\right) 
\left(\bar{u_\mu}\gamma_\lambda(1-\gamma_5)u_\nu\right)
\label{matelem}
\end{eqnarray}
where $G_F$ is the Fermi constant, $\sin\theta_c$ is the Cabibbo angle, and 
$q$, $p_K$, $p_\pi$ are the momentum transfer, the kaon and the pion 4-momenta,
 respectively.
The out-of-plane polarization ($P^T_\mu$)  of the muon is non-zero 
when the form factor $\xi$ has an imaginary component.  
This polarization is
expected to vanish to  first order 
in the standard model, because of the absence of 
the CKM phase in the decay amplitude. 
 Irreducible  
backgrounds, such as  from final-state interactions (FSI)  
in this decay are expected to be 
small ($\sim 4\times 10^{-6}$), and  can be ignored.$^{\ref{zitnitskii}, \ref{efros1}}$
It has been shown that any 
model involving only effective V or A interactions cannot 
produce this type of polarization.$^{\ref{leurer}}$ 
The existence of a non-zero
value of this polarization will therefore provide 
 a definite signature of  
physics beyond the standard  model.
$P^T_\mu$ is a function of the 
two Dalitz 
variables that define the 3-body \kmu ~decay. 
In most experimental situations, one averages 
over some portion of the Dalitz plot. Therefore, 
the average $ P^T_\mu$ has two components:
(1) a kinematic factor ($f_D$) that describes the average over
the Dalitz plot, including  experimental acceptance, and
(2) the imaginary part of an amplitude or combination
of amplitudes ($\mbox{Im}\xi$).
%  and the square of the mass of the 
%non-standard scalar intermediate particle.
We will often use the same  $P^T_\mu$ to 
denote the average over an ensemble of events:
\begin{eqnarray}
\begin{array}{ll}
 P^T_\mu = & f_D \mbox{Im}\xi  \nonumber \\
\end{array}
\label{eqn1}
\end{eqnarray}
$f_D$ is estimated by setting $\xi = 0$, 
and ignoring terms of ${\cal O} \left( {m^2_\mu\over m^2_K} \right)$:$^{\ref{garisto}}$ 
\begin{eqnarray}
f_D \approx \left\langle {m_\mu \over m_K} { \sin\theta_{\mu\nu} \over 
{E_\mu\over |\vec P_\mu| } + \cos\theta_{\mu\nu} }
      \right\rangle
\label{fd}
\end{eqnarray}
where $E_\mu$ is the muon energy and  $\theta_{\mu\nu}$ is the 
angle between the muon and the neutrino in the kaon rest
frame. The dependence
over the Dalitz plot is averaged with the experimental 
acceptance. If we choose  a typical point on the Dalitz plot, e.g.,
 $E_\mu \approx 170 MeV$ and 
$\theta_{\mu\nu} = 90^o$, then the value of 
 $f_D$ calculated at 
that point is $f_D \approx 0.17$.
The value of $\mbox{Im}\xi$ 
is model dependent;
for example, in the case of a non-standard effective scalar 
interaction, $\mbox{Im}\xi$ is 
proportional to the imaginary part of the scalar coupling strength.

In particular, multi-Higgs  and 
leptoquark models can produce non-zero out-of-plane 
 polarization. In multi-Higgs
models,  a charged Higgs particle 
mediates a scalar interaction that interferes 
with the standard model decay amplitude; in such models, 
the polarization can be as large as $10^{-3}$, without 
conflicting with other experimental constraints, 
for example,  
the measurements of the neutron electric dipole moment and 
the branching fractions for $B\to X \tau \nu$, and $b\to s \gamma$. 
The  indirect limits on $P^T_{\mu}$ from other measurements have been examined 
in the context of the minimal 3 Higgs Doublet Model 
(3HDM).$^{\ref{garisto}, \ref{belanger}, \ref{weinberg}}$
The transverse polarization for 3HDM is given by 
\begin{eqnarray}
%\begin{array}{ll}
 P^T_\mu & \approx & f_D {m^2_K \over m^2_h} { s'_\delta s'_3  c'_3 
{\nu \over \nu_1}{\nu_2 \over \nu_3} } \nonumber \\
\nu^2 &  = & \nu^2_1 + \nu^2_2 + \nu^2_3  
%\end{array}
\label{eqn2}
\end{eqnarray}
where $s'_\delta$, $s'_3$, and $c'_3$ are unknown parameters from the unitary $3\times 3$ 
mixing matrix of the 3 Higgs doublets.  
It is assumed that the ratios of the 3 vacuum 
expectation values are the same 
 as the ratios of the 3 generation masses;  $v_1:v_2:v_3::m_b:m_t:m_\tau$. 
%and $m_t\sim 174 GeV$.
 We have  updated the calculation of constraints on the 3HDM model 
 to include   new experimental results 
and collected them in Table \ref{tab_of_con}. 
 The constraints are calculated in terms of two
assumptions on the mass of 
the charged Higgs: $m_h\sim m_W$ or $m_h\sim 2 m_W$. 
Both of these assumptions are  above the 
 current lower limit of $m_h \sim 78 GeV$ 
 on the mass of a charged Higgs boson 
from LEP.$^{\ref{lephiggs}}$
 The best constraints
are from the branching ratio measurements $B(b\to s \gamma)$
and $B(b\to X \tau \nu)$. 
These are unlikely to improve in the near future because the errors
have large theoretical components. 
Furthermore, these constraints have the requirement
that the real part of the 3HDM amplitude cancels with the standard 
model amplitude. 
Without this assumption, the constraints would be 
quite weak. Nevertheless, as can be seen in the table, 
even these optimistic limits
allow any value of $P^T_\mu$ below 
the current direct bound, which is described below. 
 T-violation could occur in the minimal 
supersymmetric models through an effective scalar interaction involving the 
charged Higgs particle, nevertheless, the effect is considered too small
to observe,
except in models that contain 
R-parity violation.$^{\ref{susy1}, \ref{susy2}, \ref{susy3}}$

\begin{table}
\begin{tabular}{|l|l|ll|}
\hline
Measurement   &   Value   &   ${\bar P^T_\mu} < $ &  \\
               &         &   $m_h\sim m_W$ &  $m_h\sim 2 m_W$ \\ 
\hline 
$d_n$ [\ref{ksmith}] & $< 7.5\times 10^{-26} e-cm$ (95\%C.L.)& 0.039 & 0.039 \\
$d_e$ [\ref{abdullah}] & $< 1.6\times 10^{-26} e-cm$ (95\%C.L.)&  -- & -- \\
$\epsilon^\prime\over \epsilon$ [\ref{ktev}, \ref{na48}, \ref{aveep}] &
 $(1.9\pm 0.24)\times 10^{-3}$ & -- & -- \\
$m_{K_L} - m_{K_S}$ [\ref{pdg}, \ref{mklks}] & $(0.5301\pm 0.0014)\times 10^{10}\hbar s^{-1} $ & -- & -- \\
$B(b\to s ~\gamma)$ [\ref{alam}] & $(2.54\pm 0.56)\times 10^{-4}$ & 0.09 & 0.02 \\
$B(b\to X \tau \nu)$ [\ref{aleph}] & $(4.08\pm 0.98)\times 10^{-2}$ & 0.009 & 0.009 \\  
\hline
\end{tabular}
\caption{ \sl 
Constraints on $P^T_\mu$ of \kmu~ decay for 3HDM from  other 
measurements.$^{\protect\ref{garisto}, \protect\ref{belanger}}$  
The ``--'' means that there is no significant constraint.
All measurements have been scaled to correspond to the same  
 95\% confidence level to get consistent constraints.}
\label{tab_of_con}
\end{table}

The best limit on this process  was obtained recently  by 
an experiment  at the KEK-PS, E246.$^{\ref{kek246}}$ 
They measured  $P^T_\mu =  -0.0042 \pm 0.0049  \pm 0.0009$, which 
corresponds to a value  of the T violating parameter:
$\mbox{Im}\xi = -0.013\pm 0.016 \pm 0.003$, where the errors 
are statistical and systematic, respectively. 
This measurement was performed using approximately $3.9\times 10^6$
events.  Additional data may increase the total  sample by a factor of two.$^{\ref{kek246ichep}}$  
Experimental limits were obtained  almost 20 years 
ago with both 
neutral  and charged kaons at the BNL-AGS.$^{\ref{schmidt},\ref{campbell}}$
 The experiment with $K^+$ decays made 
a measurement of the transverse polarization, $P^T_\mu = 0.0031 \pm 0.0053$.
The combination of the neutral and charged 
 experiments could be interpreted as a limit on 
$\mbox{Im}\xi = -0.01 \pm 0.019$.

Measurements from the 1980s 
 were based on  
$1.2\times 10^7 ~K^0_L$ 
and $2.1\times 10^7 ~ K^+$ decays to $\mu^+ \pi  \nu$, and were
limited by low analyzing power  and by backgrounds.  
The KEK-E246   measurement relied on   the new technique of  
using a stopped $K^+$ beam, and 
measuring the muon decay direction in an aluminum absorber, 
without spin precession. The BNL measurements 
used in-flight kaon decays, stopped the muons in an 
aluminum absorber,  
and measured the  muon spin by its precession in a weak magnetic field. 
Both techniques relied on the cylindrical symmetry of the apparatus 
to suppress systematic errors due to non-uniform efficiencies. 
The systematic error in KEK-E246 was reduced further  by collecting events
with forward and backward going pions. The BNL experiment 
alternated the direction of the precessing magnetic field to cancel the
detection efficiency to second order. 
The remaining systematic errors in both techniques are quite 
different: the largest systematic error in the 
stopped kaon experiment comes from the knowledge of the 
fringe magnetic field in the muon stopping region.  The 
largest systematic error for the in-flight experiment is 
 thought to be due to the mechanical misalignment of the azimuthal 
segments of the muon absorber with respect to the precessing 
axis of the magnetic field.

A new experiment should reach  much higher statistics and
have high analyzing power for the muon polarization. 
KEK-246 has demonstrated that  
a well designed muon stopper with low background can have 
an analyzing power of $0.197 \pm 0.005$. 
It is possible to extend the sensitivity of the KEK-246 experiment 
with a more intense beam,$^{\ref{e936}}$ such as the low energy 
separated beam used by  BNL experiment E787,$^{\ref{lesbi3}}$ which 
has been able to obtain kaon stopping rates of approximately 1 million
per sec. A stopped kaon  experiment, with a field for precession  
has the potential advantages of having better systematic control 
using cancelations from both the forward/backward 
$\pi^0$ direction, as well as by alternating field direction.$^{\ref{e787km3}}$  

A new experiment has been designed 
at the BNL-AGS to perform this measurement,
with an error on the polarization approaching $10^{-4}$.$^{\ref{newprop}}$
 The design is based on  the 1980 experiment, however, it 
uses a   2 GeV/c separated charged kaon beam to reduce the background counting rate.
 Other improvements will involve
higher acceptance and analyzing power, 
with a larger apparatus and a more finely 
segmented  polarimeter made of graphite. 
The experiment will collect approximately 550 events per AGS 
pulse (3.6 seconds). 
Thus the statistical accuracy of the 
polarization measurement in a 2000 hr ($2\times 10^6$ pulses) run 
could be $\delta P_T \sim 1.3 \times 10^{-4}$. 
Through a precise alignment of the polarimeter and through 
measurements of null asymmetries
from $K^+ \to \mu^+ \nu$ decays
the experimenters expect to control 
the systematic error in the apparatus to below  the statistical uncertainty. 
Other studies  carried out for 
experiments at the Japanese Hadron Factory  and the $\phi$ factory at DAPHNE  
also show that 
statistical sensitivities of the order $\sim 10^{-4}$ are possible.$^{\ref{lim},
\ref{pritivera}}$

Finally, we note  that it could be easier to measure the transverse 
polarization in the case of $K_L \to \pi^- \mu^+ \nu_\mu$ decays, because 
it is easier to detect  a charged pion. 
However, the transverse polarization in this 
case is contaminated by final state interactions 
due to the presence of a charged pion. 
This effect, averaged over the phase space, 
 has recently been estimated to be 
$ P^T_\mu = 2.4\pm0.1\times 10^{-3}$.$^{\ref{efrosl}}$

\section{$K^+ \to \mu^+ \nu \gamma$}

\noindent 
The branching ratio for \rkm~ decay is $(5.5\pm 0.28)\times 10^{-3}$.$^{\ref{pdg}}$
 The decay is dominated by inner 
bremsstrahlung. Recently, the structure-dependent part of the 
branching ratio (mostly for positive-helicity photons) has been measured to be 
$(1.33\pm 0.12 \pm 0.18)\times 10^{-5}$.$^{\ref{convery}}$ 
The structure-dependent form factors have also been measured recently
in decays  of 
$K^+\rightarrow e^+\nu e^+ e^-$ and 
$K^+\rightarrow \mu^+\nu e^+e^-$, and 
 are sensitive to  the form factor for  negative-helicity 
photons.$^{\ref{e865muee}}$ 

In  \rkm~ decay,   the transverse muon polarization, which is 
T-violating, can arise from interference between  inner bremsstrahlung 
and the structure-dependent part of the decay.  
It is  sensitive to new pseudo-scalar, vector, and axial-vector  interactions, in contrast to 
\kmu~ decay's sensitivity to new scalar interactions.$^{\ref{kobayashi}}$ 
Recent studies  have shown that in certain extensions of 
the standard model, including SUSY, 
there could be  a T-violating muon polarization in \rkm~
decays as large as $10^{-2}$.$^{\ref{wu},\ref{chen}}$   
Others argue that it is unnatural to generate a contribution to 
transverse polarization in \rkm~ decay larger than $10^{-4}$ in SUSY, unless 
R-parity is broken.$^{\ref{hiller}}$ 
Finally, final-state interactions can induce  a 
non-zero transverse polarization. This has been calculated recently  
to be   of the order of 
$10^{-4}$, and varies on the Dalitz plot.$^{\ref{hiller}, \ref{efros2}}$
It can be as large as 
$5\times 10^{-4}$ at the high end of the muon energy spectrum, for  
$E_{\gamma}\sim M_K/4$.

%NEED TO CHECK ANOTHER REFERENCE BY Efrosini and Kudenko.

Following the calculations in Ref. \ref{chen}, the  \rkm~ decay is
described by 
``inner bremsstrahlung'' (IB) and ``structure-dependent'' (SD) 
contributions.  The parameters are the kaon decay constant $f_K$, 
axial vector form factor $F_A$, and a vector form factor $F_V$. 
Physics beyond the SM is introduced in additional terms in 
the Lagrangian
\begin{eqnarray}
{\cal L}&=&-{G_F \over \sqrt{2}}sin\theta_c \bar{s}\gamma^{\alpha}
(1-\gamma_5) u
\bar{\nu}\gamma_{\alpha}(1-\gamma_5)\mu+ G_S 
\bar{s}u \bar{\nu}(1+\gamma_5)\mu 
+G_P \bar{s}\gamma_5 u \bar{\nu}(1+\gamma_5)\mu \nonumber \\
&+&G_V \bar{s}\gamma^{\alpha}u
\bar{\nu}\gamma_{\alpha}(1-\gamma_5)\mu+G_A \bar{s}\gamma^{\alpha}\gamma_5 u
\bar{\nu}\gamma_{\alpha}(1-\gamma_5)\mu+h.c.,  
\label{eqn:fermi}
\end{eqnarray}
where $G_F$ is the Fermi constant, $\theta_c$ is the Cabibbo mixing angle, and
$G_S$, $G_P$, $G_V$, and $G_A$ are parameters of  the new
interactions due to  
scalar, pseudo-scalar, vector, and axial vector  exchange, 
respectively.

The new interactions will modify the decay constants and form factors in
the following way:
\begin{eqnarray}
f_K
&=& f^0_K \left(1+\Delta_P +\Delta_A\right),
\nonumber \\
 F_A&=&
F^0_A (1+\Delta_A),
\nonumber \\
 F_V&=&
F^0_V (1-\Delta_V),
\label{eqn:fk} 
\end{eqnarray}
and 
\begin{eqnarray}
\Delta_{(P,A,V)}&=&
{\sqrt{2}\over G_F sin\theta_c}
\left({G_Pm^2_K \over (m_s+m_u)m_{\mu}},G_A,G_V\right)\,.
\label{eqn:fkp}
\end{eqnarray}

\begin{figure}[p]
\begin{center}
\psfig{file=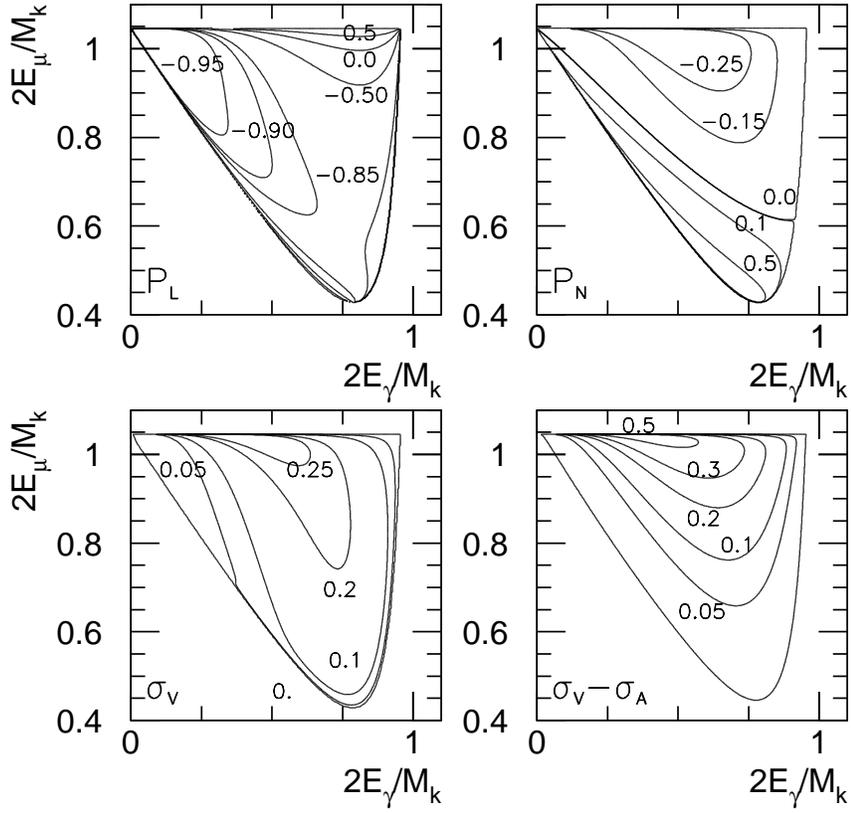,height=5in,width=5in}
\end{center}
\caption{ \it 
Muon polarization for \rkm~ decays along the 
muon momentum (top left),  normal to the muon momentum 
in the decay plane (top right), and 
$\sigma_V$ (bottom left) and $\sigma_V-\sigma_A$ (bottom right)
on the Dalitz plot.  The out-of-plane polarization would be 
proportional to the numbers on the contours 
in the bottom two plots, as given in Eq. \ref{eqn:pva}. } 
\label{rkmu2pol}
\end{figure}

The T-violating muon polarization ($P_T$) comes from the interference of IB and the
imaginary part of the SD. Only the pseudo-scalar ($G_P$) and 
right-handed current ($G_R=G_V+G_A$) terms  contribute: 
\begin{eqnarray}
P_T(x,y) &=& P^V_T(x,y)+P^A_T(x,y)
\label{eqn:pt1}
\end{eqnarray}
with
\begin{eqnarray}
\begin{array}{rcl}
P^V_T(x,y)&=&\sigma_V(x,y)[Im(\Delta_A+\Delta_V)]\,,      \\
P^A_T(x,y)&=&[\sigma_V(x,y)-\sigma_A(x,y)]Im(\Delta_P)\,,
\end{array} 
\label{eqn:pva}
\end{eqnarray}
where $\sigma_V(x,y)$ and $\sigma_A(x,y)$  are analytic functions of the 
Dalitz plot variables ($x=2E_\gamma/M_K$, $y=2E_\mu/M_K$). 
Figure \ref{rkmu2pol} shows the contours of muon
polarization along and perpendicular to the muon momentum within
the decay plane, and the contours of 
$\sigma_V$ and $\sigma_V-\sigma_A$ as a function of the Dalitz plot variables,
$E_\mu$ and $E_\gamma$.  Events with 
high muon momentum have higher sensitivity to the T-violating parameters.

The detectors designed for \kmu~ muon-polarization measurement
can be used for studying \rkm~ decays, with only minimal modification.
In the near future, the first measurement of the T-violating  polarization 
in \rkm~ is expected  from KEK-E246.  It is expected to  have 
sensitivity of the order of 5\%.$^{\ref{lim}}$ 
A \rkm~ event is usually identified in a detector through
 a coincidence of a stopped 
muon with
 an energetic acollinear photon. 
To reject  background from 
$K^+$ decays with $\pi^0$ (mainly \kmu),  events with 
additional photons are rejected.
   This requires a hermetic photon veto system surrounding 
the $K^+$ decay region.   
With a precise measurement of the photon energy and the muon momentum, 
the \rkm~ decay can be fully reconstructed.  This helps to further reduce  
background.   A precision calorimeter with good spatial resolution is
essential.  The muon momentum can be measured in a magnetic field, as in 
KEK-E246,  or the muon energy can be derived from the muon range in the 
polarimeter, as proposed in the new BNL experiment.
A study from the latter   showed that with sufficient
shielding around the decay volume, and photon veto down to about 20MeV,  
one could obtain 
a sample of $3\times10^7$ \rkm~ events, with a signal to background ratio
of about 2:1, in 2000 hours of running.$^{\ref{newprop}}$ 
 This would provide  constraints
on the new interactions at sensitivities of: 
\begin{eqnarray}
\delta_{Im(\Delta_A+\Delta_V)} &=& 7\times10^{-3} \nonumber  \\
\delta_{Im(\Delta_P)} &=& 20 \times10^{-3} 
\label{eqn:rkmsens}
\end{eqnarray}
An experiment with the sensitivity of Eq. \ref{eqn:rkmsens}, 
will clearly probe  new physics. 
Events with high muon energy have even higher
sensitivity to T-violation; they 
are also beyond the kinematic limit of \kmu~ events. They  have a 
much better signal to background ratio, but  at the expense of efficiency 
for signal. This is partially compensated by the higher sensitivity.
   For example, selecting events with high muon energy 
($2E_\mu/M_K>0.95$), 
the average $\sigma_V$  increases from 
0.11 (for all events) to 0.15; the effect is larger for    
($\sigma_V-\sigma_A$),  which increases from 0.04 to 0.25. 
We note that this is also the region where the asymmetry from FSI may be 
maximal. 
The selection of events on the Dalitz plot can be 
optimized differently for a search for pseudo-scalar or right-handed currents.

It should be noted that the T-conserving components of the muon polarization 
in \rkm~ and \kmu~ decays are of  opposite sign.    By measuring the T-conserving 
polarization of the accepted \rkm~ sample,
the background from  \kmu~ can be  easily evaluated.

\section{$K_L\to \mu^+ \mu^-$}

\noindent 
This decay has been measured  recently  with high statistics.$^{\ref{e871}}$
 Experiment E871 at BNL has collected more than 
6200 events with about 1 \% background contamination. 
The new branching ratio  is:
\begin{eqnarray}
{\Gamma(K_L \to \mu^+ \mu^-)\over \Gamma(K_L \to \pi^+ \pi^-)} = (3.474\pm 0.057)\times 10^{-6}
\end{eqnarray}
Using the measured branching ratio $B(K_L \to \pi^+ \pi^-) = (2.056\pm 0.033) \times 10^{-3}$ from
%Using the measured branching ratio $B(K_L \to \pi^+ \pi^-) = (2.067\pm 0.035) \times 10^{-3}$ from
Ref. \ref{pdg}, we obtain 
 $B(K_L \to \mu^+ \mu^-) = (7.14\pm 0.17)\times 10^{-9}$.
The phenomenology of  
this reaction has been studied extensively in the past.$^{\ref{mumurev}}$

The longitudinal muon polarization in this decay violates
CP invariance. In general, the correlation between the 
$\mu^+$ and $\mu^-$ polarizations also contains information about 
CP violation and new physics, however this is much  
more difficult to measure.$^{\ref{hemamc}}$ 
The decay amplitude for $K_L \to \mu^+ \mu^-$ is known to be dominated by 
the two photon intermediate state. Interference of this amplitude 
with any other flavor-changing neutral scalar interaction 
can produce a  longitudinal polarization. 
Following Herczeg,$^{\ref{herczeg}}$ we set the amplitude for 
$K_L \to \mu^+ \mu^-$ to:  
\begin{eqnarray}
\begin{array}{rl}
{\cal M}(K_L \to \mu^+ \mu^-) & = a \bar u(p_-)\gamma_5 v(p_+) + b u(p_-) v(p_+)   \\
a  & = a_2 + i \epsilon a_1   \\
b & = b_2 + i \epsilon b_1  \\
\end{array}
\end{eqnarray} 
where 
$a_2$ and $b_2$ are the CP conserving and CP violating amplitudes of $K_2$, respectively. 
Similarly, $a_1$ and $b_1$ are the CP violating and CP conserving 
amplitudes for $K_1$, respectively. The decay rate and the longitudinal polarization ($P_L$) are 
then given by:
\begin{eqnarray}
% \begin{array}{rl} 
\Gamma & = & {m_K \beta\over 8 \pi}(|a|^2 + \beta^2 |b|^2)   \nonumber \\ 
%			\\
P_L & = & {2 \beta \mbox{Im}(ba^*) \over |a|^2 + \beta^2 |b|^2 }  \\
%			\\
\beta & =& (1 - 4 m_\mu^2/m_K^2)^{1/2}  \approx  0.905  \nonumber 
% \end{array}
\end{eqnarray}
 
The expression for the  decay rate has several parts: 
terms  of  ${\cal O}\epsilon^2$  can be neglected, 
terms  of ${\cal O} \epsilon$ multiply amplitudes with direct 
CP violation in $K_1$ or $K_2$ and are therefore small. 
The remaining terms are $\mbox{Re}(a_2)^2 + \mbox{Im}(a_2)^2 + \mbox{Re}(b_2)^2 + \mbox{Im}(b_2)^2$.  
Each of these reflects a sum of contributions from electroweak interactions 
and possible non-electroweak physics.  The lowest-order 
standard-model electroweak contributions 
arise from the intermediate two photon diagram and the short-distance diagrams 
involving mainly the top quark, W, and Z exchange.$^{\ref{mumurev}}$ 

The largest contribution to the total branching ratio 
is known to be the absorptive contribution ($\mbox{Im}(a_2)$)
 from  two photon exchange.$^{\ref{jacksehgal}}$. This can be calculated in a model independent way:
\begin{eqnarray}
{\Gamma(K_L\to \gamma \gamma \to  \mu \mu) \over \Gamma(K_L \to \gamma \gamma)}  = 
\alpha^2 ({m_\mu \over m_K})^2 {1\over 2 \beta} {\left( \log{1+\beta \over 1-\beta} \right)}^2 = 1.195\times 10^{-5} 
\label{eqn:unit}
\end{eqnarray}
Eq. \ref{eqn:unit} is also known as the unitarity bound. All other contributions to $\mbox{Im}(a_2)$ are  
much smaller. $\mbox{Im}(b_2)$ in the standard model is due to   
$K_2 \to \pi \pi \to \gamma \gamma\mbox{(CP=+1)} \to \mu^+ \mu^-$; this is constrained to be
very small by $\epsilon'/\epsilon$. The short-distance electroweak diagrams contribute to 
$\mbox{Re}(a_2)$. The branching ratio from this 
 contribution is also well calculated in the standard model to be:
\begin{eqnarray}
% \begin{array}{cc} 
B_{SD}(K_L \to \mu^+ \mu^-) & = &   {\tau_L \over \tau_{K^+}} {\alpha^2 B(K^+ \to \mu \nu) \over 
\pi^2 \sin^4\theta_W |V_{us}|^2} \left[ Y_c \mbox{Re}(\lambda_c) + Y_t \mbox{Re}(\lambda_t) \right]^2           
% \end{array}
\end{eqnarray}
$Y_q$ are Inami-Lim functions  of $x_q = M^2_q/M^2_W$; $\lambda_j = V^*_{js}V_{jd}$ 
are  combinations of the CKM matrix elements.$^{\ref{inami}}$
Or, in terms of the Wolfenstein parameters $A$, $\rho$ and $\lambda$:
\begin{eqnarray}
B_{SD}(K_L \to \mu^+ \mu^-) 	= 1.51 \times 10^{-9} A^4 (\rho_0 - \bar\rho)^2  
\end{eqnarray}
where   $\bar\rho = \rho (1-\lambda^2/2)$, $\rho_0 = 1.2$ is the charm contribution. 

The second contribution to $\mbox{Re}(a_2)$ comes from  $K_L \to \gamma^* \gamma^* \to \mu^+ \mu^-$,
in which the photons are off mass-shell. The sign and magnitude of this contribution are still 
uncertain; this uncertainty limits the precision with which the experimental measurement 
can be used to obtain a limit on polarization,  as well as on the fundamental standard-model parameter 
$\rho$. 
First order amplitudes involving new flavor-changing vector bosons, 
leptoquarks, or  scalar particles, such as a light Higgs, contribute to $\mbox{Re}(b_2)$; 
these contributions would be responsible for a possibly large longitudinal polarization.

The unitarity bound from the absorptive amplitude 
 can be subtracted from the measured branching ratio to obtain the 
sum of all the remaining parts. In Ref. \ref{e871}, this subtraction is performed  
using the measured ratio  $\Gamma(K_L \to \mu^+ \mu^-)/ \Gamma(K_L \to \pi^+ \pi^-)$ 
and the unitarity bound calculated for the ratio
 $\Gamma(K_L \to \gamma \gamma)/ \Gamma(K_L \to \pi^+ \pi^-)$, with proper account of 
correlations in errors.$^{\ref{e791mm}}$ They obtain 
a dispersive contribution to the branching ratio of $(0.11\pm 0.18)\times 10^{-9}$:  
% We use updated numbers for the $K_L \to \gamma \gamma $ and $K_L \to \pi^+ \pi^-$
%branching ratios from [\ref{pdg}, \ref{skrev}]  to obtain the 
%dispersive contribution to the branching ratio;  
\begin{eqnarray}
{m_K \beta\over 8 \pi}( (\mbox{Re}(a^{SD}_2) \pm \mbox{Re}(a^{LD}_2))^2  + 
\beta^2 \mbox{Re}(b_2)^2) = (0.11\pm 0.18)\times 10^{-9}\times 
\Gamma(K_L)   
\label{thesum}
\end{eqnarray}
Here $\mbox{Re}(a^{SD}_2)$ is the magnitude of the amplitude due to the 
electroweak  short-distance diagrams and $\mbox{Re}(a^{LD}_2)$ is the 
magnitude of the dispersive contribution from the two photon diagram. 
We will now consider the implications of this result on the 
value of the polarization:

 Concentrating only on the direct CP violating contribution, 
the longitudinal polarization can  be written: 
\begin{eqnarray}
P_L = -{m_K \beta^2 \over 4 \pi \Gamma(K_L \to \mu^+ \mu^-)} 
\mbox{Re}(b_2)\mbox{Im}(a_2)  
\end{eqnarray} 
One can limit the magnitude of this
 polarization using the measurement in Eq. \ref{thesum}, by 
assuming $\mbox{Re}(a_2) = 0$. Then, for the square of the polarization, we can write:
\begin{eqnarray}
|P_L|^2 < 4  \times \left( 1- {\Gamma(K_L \to \gamma \gamma \to \mu^+ \mu^-)\over 
 \Gamma(K_L \to  \mu^+ \mu^-)} \right) {\Gamma(K_L \to \gamma \gamma \to \mu^+ \mu^-)\over 
 \Gamma(K_L \to  \mu^+ \mu^-)}
\end{eqnarray}
Using the Particle Data Group$^{\ref{pdg}}$ prescription, we obtain an upper 
limit\footnote[1]{Following the PDG prescription, 
 the value of $|P_L|^2$ is $0.058\pm 0.094$.  
However, it should be noted that a 
 previous result  from E791$^{\ref{e791mm}}$ has  an unphysical central value of 
$-0.06\pm 0.21$ for $|P_L|^2$.} ~for the polarization of: 
\begin{eqnarray}
|P_L| < 0.44 ~~~~~~90\% C.L. 
\end{eqnarray}
Clearly, a better limit can be obtained if $\mbox{Re}(a_2)$ can be specified 
more accurately. 

There are several mechanisms that could contribute towards generating a significant 
value of $\mbox{Re}(b_2)$ and a large polarization.$^{\ref{botella}, \ref{wolf1}, \ref{herczeg}, \ref{gengmmp}}$ 
In general, $P_L$ is quite sensitive to the presence of light scalars with CP violating 
Yukawa couplings. 
The polarization can originate  in the standard CKM model because of  
the induced $s-d-H$ coupling, where $H$ is the standard-model Higgs.
Very large polarizations can be expected for a light Higgs boson 
with a mass comparable to $K_L$. Results from LEP that limit the Higgs 
boson masses to be larger than 115 GeV,  rule out
detectable polarization from this mechanism.$^{\ref{lephiggs}}$ 
In Ref. \ref{gengmmp} multi-Higgs,  leptoquark, as well as 
left-right symmetric models are considered. Polarizations of 
the order of few percent are possible, without violating bounds from 
neutron or electron electric dipole moments, or from lepton-flavor 
violation.  Supersymmetric contributions to $P_L$ 
 are expected to be small ($\sim 10^{-3}$),  without  fine-tuning of 
parameters. 

The dominant contribution to $P_L$ within the standard model 
 is through the
 indirect CP violation parameterized by $\epsilon$.
The direct CP violating amplitudes ($b_2$ and $a_1$) can be neglected,
and using the value of $\epsilon$:$^{\ref{winwolf},\ref{pdg}}$
\begin{eqnarray}
\epsilon &=& 2.3\times 10^{-3} e^{i\phi} \\
\phi &=& 43.7^o\pm 0.1^o \approx {\pi\over 4} 
\end{eqnarray}
The polarization can be written as 
\begin{eqnarray}
% \begin{array}{rcll}
P_L &=& {m_K \beta^2\over 4 \pi \Gamma(K_L \to \mu^+ \mu^-) \sqrt{2}} |\epsilon|
  [\mbox{Re}(a_2)
(\mbox{Re}(b_1) -\mbox{Im}(b_1))  \nonumber \\ 
&   +&  \mbox{Im}(a_2)(\mbox{Re}(b_1) +\mbox{Im}(b_1))] 
% \end{array}
\end{eqnarray}
Ecker and Pich  have calculated the value of $b_1$
using chiral perturbation theory (CHPT),  to obtain a rather good 
prediction for $P_L$.$^{\ref{epichmumu}}$   They obtain:
\begin{eqnarray}
(1.9) 1.5 < |P_L| \cdot 10^3 \sqrt{2\times 10^{-6}\over B(K_S \to \gamma \gamma)} < 2.5 (2.6)  
\end{eqnarray}
The numbers without (with) brackets correspond to 1 standard deviation, $\sigma$, (2 $\sigma$) errors. There are three 
sources of uncertainty in the above estimate. The uncertainty on the octet coupling strength ($G_8$)
in CHPT is largely eliminated by normalizing to 
the measured $B(K_S\to \gamma \gamma) = (2.4\pm 0.9) \times 10^{-6}$.$^{\ref{ksgg}}$
 The uncertainty 
in the upper limit is small because it 
comes from $\mbox{Im} (a_2)$. The uncertainty in the lower limit comes from the unknown sign and 
magnitude of $\mbox{Re}(a_2)$. Ecker and Pich point out that there is a constructive interference 
in the term multiplying $\mbox{Im}(a_2)$,
 which makes their estimate larger than a previous estimate.$^{\ref{herczeg}}$
 To summarize,  the polarization due to indirect CP violation is
estimated to be  $\sim 2 \times 10^{-3}$
with  30-40\% uncertainty.

The main experimental difficulty in this measurement is the 
small branching fraction of  $(7.14\pm 0.17)\times 10^{-9}$ for the decay.
Much effort must therefore be put into separating these 
events from background, before a polarization analysis can be 
performed. 
 Experiment E871,$^{\ref{e871}}$ with $\sim 6200$ events,  was 
optimized to look for $K_L \to \mu^\pm e^\mp$.
 We have  estimated that,  
if the experiment  were optimized for 
$K_L\to \mu^+ \mu^-$, and the beam intensity were increased,
E871 could collect about 20,000 events in two 
years of running.  Appropriate upgrades to the marble 
muon-range detector will allow approximately 50\% of the positive 
muon decays 
to be analyzed.
A  study by the E791 collaboration (previous version of E871),
indicated that, with 20,000  events, they could achieve a sensitivity 
 of 11\% on the longitudinal polarization asymmetry
of the positive muons.$^{\ref{e791mumu}, \ref{mumupol}}$

The $\sim 1\%$ background to $K_L \to \mu^+ \mu^-$
 arises  mainly from $K_L \to \pi^\pm \mu^\mp \nu$ events, in which 
the charged pion decays in flight or is misidentified as a muon, and 
the momentum of one of the charged particles is mismeasured so that 
the $\mu \mu $ invariant  mass is  
 higher than the kinematic limit of 489 MeV. 
The positive muons in the  $K_L \to \pi^- \mu^+ \nu$ 
background events
at the kinematic endpoint will be almost completely longitudinally  polarized; 
the positive muons in the background $K_L \to \pi^+ \mu^- \nu$ will come from 
$\pi^+$ decay, and will be polarized (with strength dependent on 
the $\pi^+$ decay angle within the experimental acceptance)
 in the opposite direction. 
Given these considerations, this background will not be problematic 
for an experiment with a sensitivity of $\sim 0.1$, however,  
a measurement  of the $K_L \to \mu^+ \mu^-$ 
polarization  better than 1\% will require
 less background.

\section{$K^+\to \pi^+ \mu^+ \mu^-$}

\noindent 
This decay has a very rich structure that can
 lead to  important measurements. Table 
\ref{list1} shows three 
different  asymmetries that could be interesting to measure:
longitudinal muon polarization, transverse muon polarization, and 
transverse $\mu^\pm$ polarization in correlation with  $\mu^\mp$
longitudinal polarization.
The decay has recently been analyzed  extensively.$^{\ref{wise1}-\ref{anbg}}$
The  different processes that govern the decay are:
the one-photon intermediate state, a two-photon intermediate
state, the short-distance ``Z-penguin'' and ``W-box'' graphs, and
potential contributions from extensions to the Higgs sector.

The dominant amplitude from the one photon intermediate state (with a vector form factor)
 is best understood in the framework of 
chiral perturbation theory, although experimental inputs are needed to fully 
describe the decay.$^{\ref{dambrosio}}$ 
The $K^+ \to \pi^+ l^+ l^-$ decays can be discussed  using the following 
variables:  
\begin{eqnarray} 
x &  = &  {(p_1 + p_2)^2\over m^2_K} \nonumber \\
y & =  & {2 P\cdot(p_1 -p_2) \over m^2_K \lambda^{1/2}(1,x,x_\pi)} \nonumber  \\
  & = &  {2 k \cdot(p_1 - p_2) \over m^2_K \lambda^{1/2}(1,x,x_\pi)} 
\end{eqnarray}
where $p_1$ and $p_2$ are the four-momenta of the negative and positive 
lepton. $P$ and $k$ are the four-momenta of the kaon and the pion, respectively. 
$\lambda(a,b,c) = a^2 + b^2 + c^2 - 2(ab + bc + ac)$ and $x_\pi = m^2_\pi/ m^2_K$.  
The limits of the phase space are given by:
\begin{eqnarray} 
{4m^2_l\over m^2_K}  < &  x & < {(m_K - m_\pi)^2 \over m^2_K} \nonumber \\  
|y| & < & (1 - {4 m^2_l\over x m^2_K})^{1/2} 
\end{eqnarray}  
In the kaon rest frame, the energy of the pion is given by: $E_\pi = (M_K/2)(1-x-x_\pi)$.
In terms of these variables, the decay rate is given by 
\begin{eqnarray} 
{d\Gamma \over dx dy}  & = & {m_K \over 64 (2\pi)^3 } \lambda^{1/2}(1,x,x_\pi) |{\cal M}|^2 
\end{eqnarray}  
where $\cal{M}$ is the matrix element. 
For a vector interaction model, the decay rate is given by the following:$^{\ref{ecker291},\ref{greenlee}}$  
\begin{eqnarray} 
{d\Gamma \over dx dy}  & = & {\alpha^2 |G_8|^2 m^5_K \over 16 \pi } \lambda^{3/2}(1,x,x_\pi) (1-y^2) |\phi(x)|^2  
\end{eqnarray}  
where $G_8$ is the octet coupling constant.
The general features of $K \to \pi l^+ l^-$ decay spectrum can be understood 
in terms of angular momentum conservation.
 In the vector interaction model, the quantum numbers of the 
$l^+ l^-$ pair are $J^{pc} = 1^{--}$. 
The lepton  pair must therefore 
 be  in a p state relative to the pion, and 
it must be longitudinally polarized. 
This has two consequences: the angle between 
the lepton and the pion in the rest frame of the lepton pair has a distribution 
that goes as $(1-\cos^2\theta)$ (This is the reason for the $(1-y^2)$ term; 
$y\approx \cos \theta$,
 is  the difference in the
energies  of the lepton pair divided by the pion momentum 
 in the kaon rest frame);
 as the mass of the lepton pair gets large, 
the orbital angular momentum barrier forces the decay spectrum to 
fall faster than phase space, and to vanish at the kinematic endpoint.

The form factor $\phi(x)$ has been measured very well by experiment E865,
with more than 10,000 events in the   $K^+\to\pi^+e^+e^-$ decay mode.$^{\ref{e865pee}}$
The form factor is  linear in $x$, with a large  
 slope parameter that has been measured  with high precision.
The dominance of the
vector current in the decay has been demonstrated experimentally. 
The measurement puts significant constraints on models containing  
long-distance contributions to $K \to \pi l^+ l^-$ decays. 
The  $K^+\to \pi^+ \mu^+ \mu^-$ decays 
have  been  measured 
in two different experiments, $\sim 200$ events in E787,$^{\ref{e787pmm}}$
 and $\sim 400$ events 
in E865.$^{\ref{e865pmm}}$ 
The average branching ratio is $7.6\times10^{-8}$, but the two experiments
disagree  by 3.3$\sigma$. 
\begin{eqnarray}
B(K^+\to\pi^+\mu^+\mu^-, {\rm E787})
         &=& (5.0\pm0.4(stat)\pm0.7(syst)\pm0.6(th))\times 10^{-8} ;  
\nonumber \\
B(K^+\to\pi^+\mu^+\mu^-, {\rm E865})
                         &=&(9.22\pm0.60(stat)\pm0.49(syst)) \times 10^{-8}.
\end{eqnarray}
Using the more recent  $K^+ \to \pi^+ e^+ e^-$ measurement of the form factor, 
the $K^+ \to \pi^+ \mu^+ \mu^-$ branching ratio is predicted to be 
$(8.7\pm 0.4)\times 10^{-8}$.
The E787 result  used an old value for the  form factor; the theoretical error 
   in their measurement corresponds to this form factor.
The disagreement between the two measurements is large even 
 if we  correct  the E787 result using the new form factor.

In general,  the decay process can have contributions from 
scalar, vector, pseudo-scalar and 
axial-vector interactions, with corresponding form factors, $F_S$, $F_V$, 
$F_P$, and $F_A$ (following the notation in Ref. \ref{anbg}):
\begin{eqnarray}
{\cal M} & = & F_S \bar u(p_l, s)v(\bar p_l, \bar s) 
+ F_P \bar u(p_l, s)\gamma_5 v(\bar p_l, \bar s)  \nonumber \\ 
 & & + F_V p^\mu_k \bar u(p_l, s)\gamma_\mu  v(\bar p_l, \bar s) 
    + F_A p^\mu_k \bar u(p_l, s)\gamma_\mu  \gamma_5 v(\bar p_l, \bar s) 
\label{amppmm}
\end{eqnarray}
Here $p_k$, $p_\pi$, $p_l$, and $\bar p_l$ are the kaon, pion, lepton, and 
antilepton 4-momenta. 
Any interference between the terms with  a complex phase difference
leads to polarization effects. 
Within the standard model, 
the largest contribution is from the one-photon intermediate state to the 
vector form factor ($F_V$), which is expected to be almost 
real.  The scalar form factor is expected to get 
a small contribution from only the two-photon intermediate state.
  $F_P$ and $F_A$ get contributions from the 
short-distance ``Z-penguin'' and ``W-box'' diagrams,
 where the dominant term arises 
from t-quark exchange,  both form factors are therefore proportional to 
$V_{ts}V^*_{td}$, with a small contribution from the charm quark.  

\begin{figure}
\begin{center}
\psfig{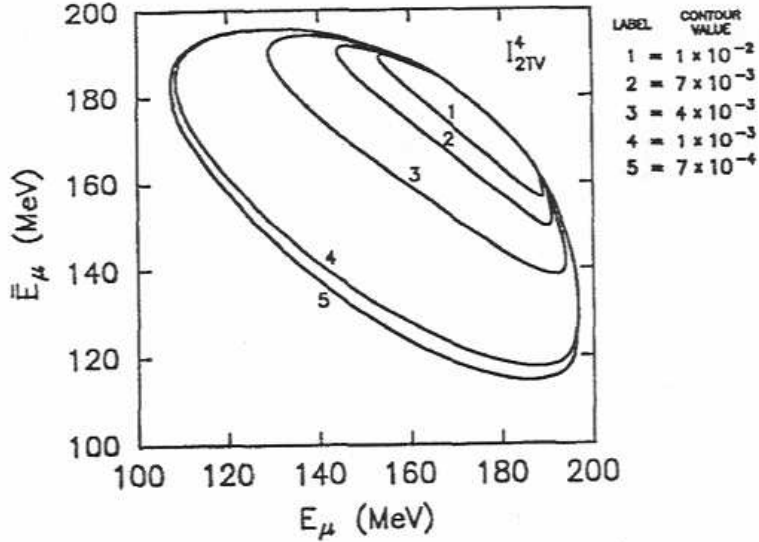}
\end{center}
\caption{\it T-violating spin-spin correlation, or asymmetry (6)
  in Table \ref{list1}, for 
$K^+ \to \pi^+ \mu^+ \mu^-$ over the Dalitz plot. 
(After Agrawal, Ng, Belanger, and Geng$^{\ref{anbg}}$).}
\label{pmmass}
\end{figure}

The  parity-violating  
longitudinal muon polarization (asymmetry (4) in Table \ref{list1}) has terms 
  proportional to  $\mbox{Re}(F_PF^*_V)$ and to $\mbox{Re}(F_VF^*_A)$. 
It is therefore  sensitive to the  Wolfenstein parameter 
 $\rho$. The value of this polarization within the standard model 
is estimated beyond the leading logarithms 
to be $ 0.003 - 0.0096$, for $-0.25 \le \rho \le 0.25$, 
and depends on the experimentally 
accessible region of phase space.$^{\ref{buchalla}}$
Neglecting the dependence of the form factor, the polarization 
asymmetry varies as $(1- 4 m^2_\mu/x  m^2_k) \lambda^{3/2}(1, x, x_\pi)$.
It has a maximum at a $\mu \mu$ invariant mass
of approximately 250 $MeV/c^2$.
% and drops to zero 
%as the muon-muon mass increases to the kinematic limit. 
There is a  non-negligible
contribution to this 
polarization from the long-distance two photon process, which has not 
as yet been calculated accurately.$^{\ref{wise1}}$ 
The T-violating transverse polarization (asymmetry (5) in Table \ref{list1}) 
is proportional to  $\mbox{Im}(F_S F_V^*)$, and it is therefore  
 expected to be  small within the standard model, and the 
final-state interaction correction to this polarization is expected to 
be  $\sim 10^{-3}$.
 The T-violating spin-spin correlation that involves 
both $\mu^+$ and 
$\mu^-$ polarizations (asymmetry (6) in Table \ref{list1})
has terms  proportional to  $-\mbox{Im}(F^*_V F_P)$
 and $\mbox{Im}(F_V F^*_A)$. 
This  is expected to have much smaller final-state 
interaction corrections, and is theoretically sound. 
Within the standard model, 
such asymmetry is proportional to the CKM parameter $\eta$, and 
it is expected to increase with  $\mu \mu$ invariant mass,
becoming as large as $\sim 0.06$ in some parts of the decay phase space
(see Fig. \ref{pmmass}). 
  A good measurement would certainly be valuable in 
understanding CP violation.
Similar to $K_L \to \mu^+ \mu^-$,
this asymmetry could also receive contributions from
non-standard physics, such as 
from charged Higgs or leptoquarks.$^{\ref{anbg}}$ 
The  consensus  in the literature on the numerical results 
is summarized  in Table \ref{list2}. It should be noted that 
 these estimates were made before the precise measurement 
of the form factor $F_V$ in Ref. \ref{e865pee}, nevertheless,
that should not affect 
substantially the values in Table \ref{list2}.

The main experimental difficulty is in isolating
the rare  $K^+\to \pi^+ \mu^+ \mu^-$ decays from background. 
For both E787 and E865, the main 
background is the \kpp ~decay, which has a branching ratio 6 orders of 
magnitude higher.
This background must be suppressed at  the  trigger level as well as 
in the final analysis. 
In E865, the charged pions can be misidentified as muons due to pion decays
in flight. In E787, where the unique total kinetic energy is used as the 
signature,  the additional energy release in $\pi^-$ nuclear capture
can increase the apparent kinetic energy of \kpp~ events. 
Neither experiment was optimized for $K^+\to \pi^+ \mu^+ \mu^-$, and  
both experiments  took dedicated 
 $K^+\to \pi^+ \mu^+ \mu^-$ triggers in  only a limited 
running period.  

Scaled from E865, a dedicated and optimized experiment, with 
improved acceptance and trigger capability for muons, increased kaon flux with 
reduced beam halo, longer running time, and an excellent muon polarimeter, 
could  yield $\sim 50,000$ events.
This could result in a measurement of $\mu^+$ polarization with 
$\delta P_L\sim 0.05$, assuming the same analyzing power as
the estimate for the polarization measurement 
of $K_L\to \mu^+ \mu^-$. 
An interesting option for this measurement could be to rely on 
 the decay-at-rest technique used by E787: 
 the \kpp ~background could be  suppressed by selecting 
the $\pi^+$ momentum 
to be above the kinematic limit of \kpp ~decays (125 MeV/c); then both 
$\mu^+$ and $\mu^-$ would have low kinetic energies, and could be stopped and
analyzed  in the  same active target that  serves to stop  incoming kaons. 
Clearly, measuring asymmetries that require analyzing both $\mu^+$
and $\mu^-$ polarizations will be difficult 
because  $\mu^-$ 
decays have a  lower analyzing power due to muon capture 
into atomic orbits around nuclei in the muon stopper. 

Authors of
Ref. \ref{anbg} point out that  
there are only a few interesting kaon measurements that 
offer clean theoretical interpretation in terms of the standard 
 weak-interaction parameters. The measurement of 
the branching ratio of 
$K^+ \to \pi^+ \nu \bar\nu$ provides a measurement of 
$|V^*_{ts} V_{td}|$, the branching ratio  $K^0_L \to \pi^0 \nu \bar\nu$ is proportional 
to $\eta^2$.$^{\ref{e787pnn}-\ref{littpnn}}$
 The measurement of the polarization asymmetries (4) and (6) from Table \ref{list1}   
%in $K^+ \to \pi^+ \mu^+ \mu^-$ decays 
could be of comparable significance 
in terms of the overall theoretical uncertainties. 
The experimental difficulty 
should be similar 
 to other proposed  investigations of  
the K and B systems.

%\vspace{1cm} 

\section{$K^0_L \to \pi^0 \mu^+ \mu^-$}

\noindent 
The structure of 
 $K_L\to \pi^0 l^+ l^-$ decays is more complex than 
 that of  $K^+ \rightarrow \pi^+ l^+ l^-$ decays 
because of  CP suppression.  
 There are three possible 
contributions to the decay amplitude: 1) direct 
CP-violating contribution from electroweak penguin
and W-box diagrams, 2) indirect CP-violating amplitude 
from the $K_1 \to \pi^0 l^+ l^-$ component in $K_L$, and 
3) CP-conserving amplitude from the $\pi^0 \gamma \gamma$
intermediate state. The sizes of the three contributions 
depend on the final-state lepton.$^{\ref{winwolf}, \ref{rwoj}}$
The CP-conserving two-photon contribution to the electron mode is expected to 
be $(1-4)\times10^{-12}$, based on $K_L\rightarrow \pi^0 \gamma \gamma$ data.
Although suppressed in phase space, this contribution to the muon mode is 
comparable to the electron mode because of a term proportional to 
lepton mass.$^{\ref{hsehgal}}$   The  CP-violating contribution to 
the electron mode is expected to be
\begin{eqnarray}
B(K_L \to \pi^0 e^+ e^-)_{CPV} =
  \left[ 15.3 a_S^2 - 6.8 { {Im \lambda_t} \over 10^{-4}} a_S
           + 2.8 \left({Im \lambda_t}\over {10^{-4}}\right)^2 \right] \times 10^{-12}
\end{eqnarray}
and the muon mode is suppressed by a factor of 5 due to phase space.$^{\ref{dambrosio}}$  
$a_S$ is an unknown parameter in the 
 $K_S\rightarrow\pi^0l^+l^-$ vector form factor, and $\lambda_t = V^*_{ts}V_{td}$.

%The standard model expectation is $B(K_L \to \pi^0 \mu^+ \mu^-) \sim (0.44-1.00)\times 10^{-11}$
%with the CP violating contribution of $\sim 0.8\times 10^{-12}$.$^{\ref{pi0mumu}, \ref{hsehgal}}$. 

The modes $K_L \to \pi^0 e^+ e^-$ and $K_L \to \pi^0 \mu^+ \mu^-$ 
  have not as yet been  observed; 
the  current  best limits on the branching ratios  for $K_L \to \pi^0 l^+ l^-$
 were obtained by the KTeV experiment 
at FNAL;  
$B(K_L \to \pi^0 \mu^+ \mu^-) < 3.8 \times 10^{-10}$
and $B(K_L \to \pi^0 e^+ e^-) < 5.1\times 10^{-10}$.$^{\ref{ktevpmm}, \ref{ktevpee}}$
These limits were based on  2 observed events in each case, 
 and expected backgrounds  of
$0.87\pm 0.15$ for the muon mode and $1.06\pm 0.41$ for the electron mode. 

The main backgrounds for the muon mode 
 were estimated to be from $\mu^+ \mu^- \gamma \gamma$ 
($0.37\pm 0.03$)  and $\pi^+ \pi^- \pi^0$ ($0.25\pm 0.09$), in which both charged pions 
decay in flight.  Of these, the former background could be irreducible and therefore of 
great concern. 
The decay $K_L \to \mu^+ \mu^- \gamma \gamma$ proceeds via the
 Dalitz decay $K_L \to \mu^+ \mu^- \gamma$, with 
an internal bremsstrahlung photon. 
KTeV has detected 4 such events, with expected background of $0.16\pm 0.08$ resulting in a 
branching ratio of $B(K_L \to \mu^+ \mu^- \gamma \gamma) = 
(10.4^{+7.5}_{-5.9}\pm 0.7) \times 10^{-9}$, consistent with the 
expectation of $(9.1\pm 0.8)\times 10^{-9}$ obtained from QED  and the measurement of 
$K_L \to \mu^+ \mu^- \gamma$.$^{\ref{ktevmmgg}}$   
Such decays were analyzed by Greenlee as background to 
$K_L \to \pi^0 e^+ e^-$.$^{\ref{greenlee}}$
He showed that the background from $e^+ e^- \gamma \gamma $ 
can be suppressed by removing events with
 photon-photon invariant masses near the $\pi^0$ mass, and
by constraining  the energies and angles of accepted  photons. 
Unfortunately, the  criteria 
 that can be used for suppressing $e^+ e^- \gamma \gamma$ are 
not effective for 
 the $\mu^+ \mu^- \gamma \gamma$ channel
 because the invariant mass of the photons is 
 restricted by the  energy 
available
 in the latter decay.  
In KTeV,
 the $\pi^0$ mass resolution was expected to be 2.4 MeV; 
they applied a 2.5 $\sigma$
 cut around the $\pi^0$ mass 
to obtain the estimated background of 
$0.37\pm 0.03$ from $\mu^+ \mu^- \gamma \gamma$. 
For any
 future experiment wishing to 
 measure the
 polarization of the muon, a good range measurement of the 
muons will be required, which
 will most likely suppress the other backgrounds 
found 
in KTeV ($\pi^+ \pi^- \pi^0$ with pion decay to muons, and $\pi^\pm \mu^\mp \nu$ with an 
accidental $\pi^0$). 
However, it seems unlikely that the 
background due to $\mu^+ \mu^- \gamma \gamma$ 
can be lowered. 
The single-event sensitivity of the KTeV result is quoted as $7 \times 10^{-11}$;  
the signal to background ratio, assuming 
that only $\mu^+ \mu^- \gamma \gamma$ will contribute in a future experiment, 
will therefore be around 1/5, if the standard-model signal is taken as  
$B(K_L \to \pi^0 \mu^+ \mu^-) \sim 5\times 10^{-12}$.

\begin{figure}[p]
\begin{center}
\psfig{file=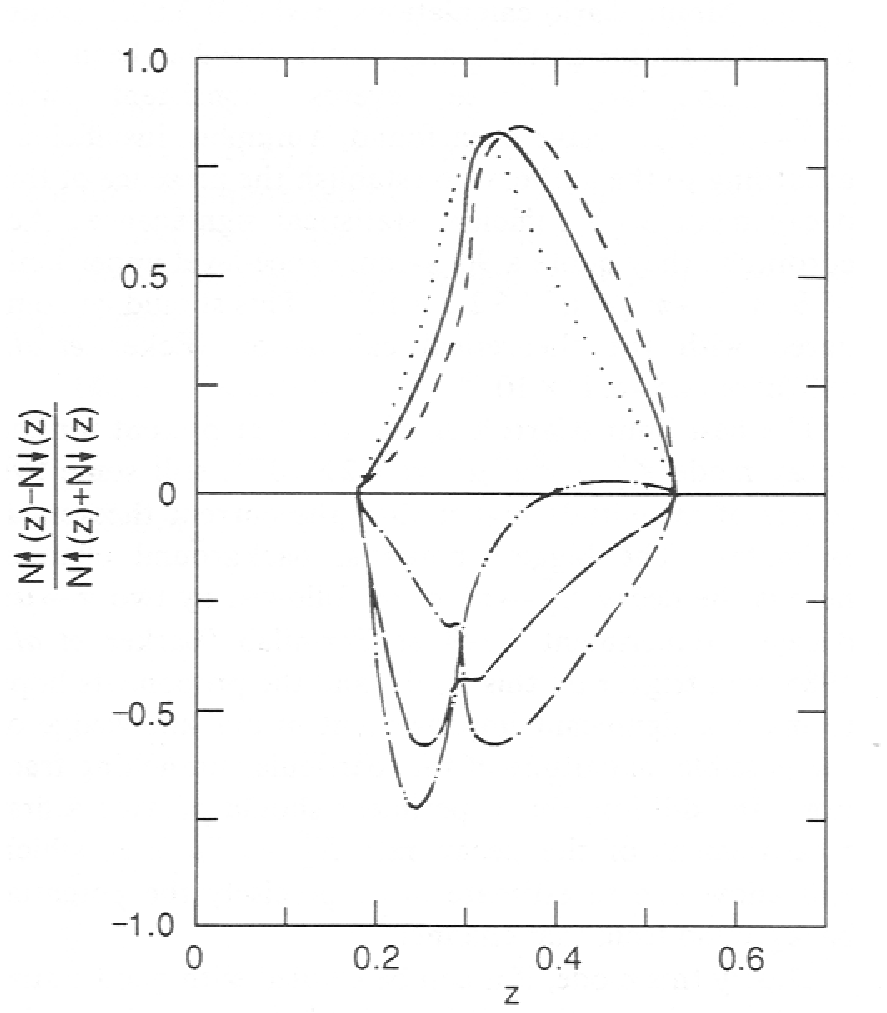,height=3in,width=3in}
\end{center}
\caption{ \it 
Up-down asymmetry in $K_L \to \pi^0 \mu^+ \mu^-$ 
decay as a function of $z$ (defined as $z= m^2_{\mu\mu}/m^2_K$),
for two possible values of $\mbox{Re}(w_s)$ and three different 
values of $\mbox{Im}(w_s)$ covering the expected ranges for 
this parameter. 
 The different curves correspond to the 
following values of ($\mbox{Re}w_s$, $\mbox{Im}w_s$):
double-dot-dashed (0.73, $-10^{-3}$), long-dashed (0.73, 0), 
dot-dashed (0.73, $+10^{-3}$), dashed (-1.00, $-10^{-3}$), 
solid (-1.00,0), dotted (-1.00,$+10^{-3}$), (after Ecker, Pich, 
and de Rafael.$^{\ref{pi0mumu}}$) } 
\label{pmmfig}
\end{figure}

The interesting direct CP-violating component 
must be extracted from any signal found 
for $K_L \to \pi^0 l^+ l^-$ in the presence of two formidable obstacles:
1) the theoretical uncertainty on contamination from indirect CP-violating and 
CP-conserving contributions and 
2) the experimental background from the $l^+ l^- \gamma \gamma$. 
A complete discussion 
 of this situation is beyond the scope of this article. In other 
reviews,  it is stated that,
 in the case of $K_L \to \pi^0 e^+ e^-$ 
 other measurements may be needed to understand the direct CP-violating 
contribution: 
e.g., the branching  ratio for
 $K_S \to \pi^0 e^+ e^-$ and the energy asymmetry between $e^+$ and
 $e^-$.$^{\ref{rwoj},\ref{bgeng91}}$
The branching ratio measurement of the $\pi \mu \mu$ mode 
provides additional information, such as 
an alternative measurement of the contribution due 
to the single-photon and two-photon intermediate states. 
The measurement of muon polarization in the $\pi \mu \mu$ 
case could also provide 
important additional information. 
This can be seen as follows: the decay amplitude can be divided 
into scalar (S), pseudo-scalar (P), vector (V), and axial-vector (A) parts. 
Of these, 
 the  scalar piece,
 which comes from the two-photon intermediate 
state is CP-conserving and  the indirect 
CP-violating contribution is mostly vector.  
The short-distance direct CP violating amplitude has both
 vector and axial-vector parts.
The longitudinal muon polarization
(asymmetry (7) in Table \ref{list1}), while not strictly CP-violating,
 can arise only  from 
interference of P and A  with S and V, and therefore it provides 
information on  
the direct CP-violating amplitudes.   
Unlike for $K^+$, in the case of $K_L$ decay,  
all the amplitudes should be of the same order of magnitude, and  
 polarization effects should therefore be very large ($\sim 1$), unless there
are strong cancelations. The interference of scalar and 
vector components can give rise to transverse muon polarization (asymmetry (8)
in Table \ref{list1}).  
This was studied in the context of a CHPT calculation of $O(p^4)$.$^{\ref{pi0mumu}}$
The transverse polarization  is displayed in Fig. \ref{pmmfig}
as a function of      $z= m^2_{\mu\mu}/m^2_K$, for several values of 
the renormalization parameter $w_s$.
In this calculation, $\mbox{Im}(w_s)$ quantifies the direct CP-violation in the 
vector part of the matrix element.  
At the time of that study, the parameter $\mbox{Re}(w_s)$ was thought to be 
constrained by the measurement of $K^+\to\pi^+e^+e^-$. Later, it was 
 argued that 
the constraint is model dependent, 
and $\mbox{Re}(w_s)$
 should be considered as unknown as $a_S$.$^{\ref{donoghue},\ref{dambrosio}}$ 
Although a new detailed calculation is needed, especially to incorporate
the new knowledge of the CP-conserving contribution,  
it is clear that the transverse polarization can be large.

These large asymmetries should be easy to measure with sufficient
statistics.  This is no longer out of question, if one considers that the
proponents of the BNL
experiment KOPIO expect to measure $\sim 50$ 
$K_L \to \pi^0 \nu \bar \nu$ events.$^{\ref{e926}}$ 
Measuring the muon polarization asymmetries in $K_L \to \pi^0 \mu^+ \mu^-$,
together with the branching ratio and the lepton energy asymmetry,  
could be a good way of defeating  the intrinsic background from 
CP-conserving and indirect CP-violating amplitudes and
  the experimental background from $\mu^+ \mu^- \gamma \gamma$. 

\section{Conclusion}

\noindent 
Muon polarization from kaon decays has a rich phenomenology.
In the case of $K_L \to \mu^+ \mu^-$,  
$K^+ \to \pi^+ \mu^+ \mu^-$ and $K_L \to \pi^0 \mu^+ \mu^-$, 
new measurements could lead to important constraints on 
the  CKM parameters, in particular 
the Wolfenstein parameters $\rho$ and $\eta$. 
The experimental  difficulties should be comparable
 to those
facing the rare kaon decay measurements of 
$K^+ \to \pi^+ \nu \bar\nu$ 
and $K_L \to \pi^0 \nu \bar\nu$,
which  are considered the best
 modes for understanding 
short distance physics in the kaon sector  (for a recent review, 
see Ref. \ref{skrev}). 
In particular, the flux available for the new experiment E926 for $K_L
\to \pi^0 \nu \bar \nu$,
could be sufficient for a measurement of polarization 
in $K_L \to \pi^0 \mu^+ \mu^-$.$^{\ref{e926}}$ 
As shown in Tables 
\ref{list1} and \ref{list2},  
for many cases, 
limits on the muon polarization will probe new physics 
beyond the standard model. In particular, 
the polarization will be sensitive to the physics of
a more complicated Higgs sector,
 or leptoquarks, that could 
give rise to CP or T violation from sources 
 outside  of the standard model.
% CP-violation needed for baryogenesis could be  
%the motivation for such searches. 

We have examined the measurement of the out-of-plane 
muon polarization in \kmu ~and \rkm ~decays. Such 
measurements will not be sensitive to sources of 
CP violation in 
the standard model. 
Nevertheless, the measurements can be 
performed with  sensitivity approaching $\delta P \sim 10^{-4}$
for \kmu,  ~and $\delta P \sim 10^{-3}$ for \rkm.  
For \kmu ~decays, this 
is well beyond  the current direct limit 
of $(-4.2 \pm 4.9)\times 10^{-3}$, and  the indirect 
 limit of $\sim 10^{-3}$, available  from 
other experimental constraints.
Although the electric dipole moments  of the
neutron and electron are considered more favorably
for T violation outside the standard model, they do not 
cover the entire  spectrum of possibilities 
beyond the standard model. 
At the moment, the measurement of T-violating 
polarization in \kmu ~and \rkm  ~decays 
is well justified and should be considered complementary 
to other efforts in understanding CP violation.

We  thank Laurence Littenberg, Larry Trueman, 
Yoshi Kuno, Steve Kettell for useful discussions.  This work was 
supported by DOE grant DE-AC02-98CH10886.

\end{document}